\documentclass[pra,twocolumn,amsmath,amssymb,aps,superscriptaddress]{revtex4-2}
\usepackage{graphicx}% Include figure files
\usepackage{color}
\usepackage{amsfonts}
\usepackage{physics}
\usepackage[capitalise]{cleveref}

\begin{document}
\title{Noise Protected Logical Qubit in a Open Chain of Superconducting Qubits with Ultrastrong Interactions}
\author{Roberto Stassi} \email{rstassi@unime.it}
\affiliation{Dipartimento di Scienze Matematiche e Informatiche,Scienze Fisiche e Scienze della Terra, Universit\`a di Messina, I-98166 Messina, Italy}
\affiliation{Theoretical Quantum Physics Laboratory, RIKEN, Wakoshi, Saitama 351-0198, Japan}
\author{Shilan Abo}
\affiliation{Institute of
Spintronics and Quantum Information, Faculty of Physics, Adam Mickiewicz University, 61-614 Poznan, Poland.}
\author{Daniele Lamberto}
\affiliation{Dipartimento di Scienze Matematiche e Informatiche,Scienze Fisiche e Scienze della Terra, Universit\`a di Messina, I-98166 Messina, Italy}
\author{Ye-Hong Chen}
\affiliation{Fujian Key Laboratory of Quantum Information and Quantum Optics, Fuzhou University, Fuzhou 350116, China.}
\affiliation{Theoretical Quantum Physics Laboratory, RIKEN, Wakoshi, Saitama 351-0198, Japan}
\author{Adam Miranowicz}
\affiliation{Institute of
Spintronics and Quantum Information, Faculty of Physics, Adam Mickiewicz University, 61-614 Poznan, Poland.}
\affiliation{Theoretical Quantum Physics Laboratory, RIKEN, Wakoshi, Saitama 351-0198, Japan}
\author{Salvatore Savasta}
\affiliation{Dipartimento di Scienze Matematiche e Informatiche,Scienze Fisiche e Scienze della Terra, Universit\`a di Messina, I-98166 Messina, Italy}
\author{Franco Nori}
\affiliation{Center for Quantum Computing, RIKEN, Wakoshi, Saitama 351--0198, Japan}
\affiliation{Physics Department, The University of Michigan, Ann Arbor, Michigan 48109-1040, USA.}

\date{\today}

\begin{abstract}

To achieve a fault-tolerant quantum computer, it is crucial to increase the coherence time of quantum bits. In this work, we theoretically investigate a system consisting of a series of superconducting qubits that alternate between XX and YY ultrastrong interactions. 
By considering the two-lowest energy eigenstates of this system as a {\it logical} qubit, we demonstrate that its coherence is significantly enhanced: both its pure dephasing and relaxation times are extended beyond those of individual {\it physical} qubits.
 Specifically, we show that by increasing either the interaction strength or the number of physical qubits in the chain, the logical qubit's pure dephasing rate is suppressed to zero, and its relaxation rate is reduced to half the relaxation rate of a single physical qubit.
Single qubit and two-qubit gates can be performed with a high fidelity.

\end{abstract}
\maketitle

\begin{figure}[b!]
    \centering
    \includegraphics[width=0.95\columnwidth]{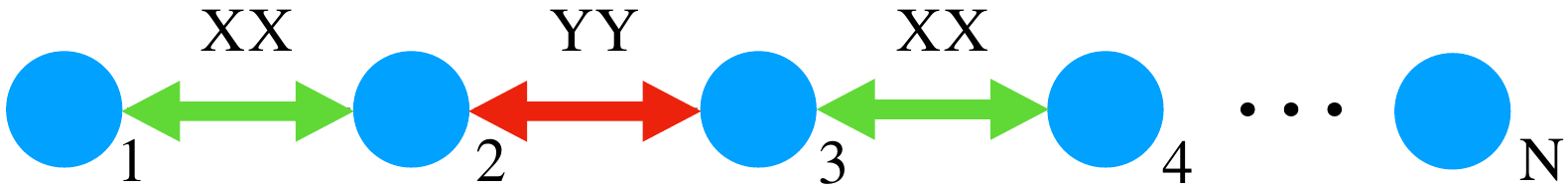}  % Adjust width as needed
    \caption{Sketch of an open chain of ultrastrongly coupled atoms in an alternating XX and YY configuration.}
    \label{Fig1}
\end{figure}

\section{INTRODUCTION}

\begin{figure*}[t]
\centering
  \includegraphics[scale=0.5]{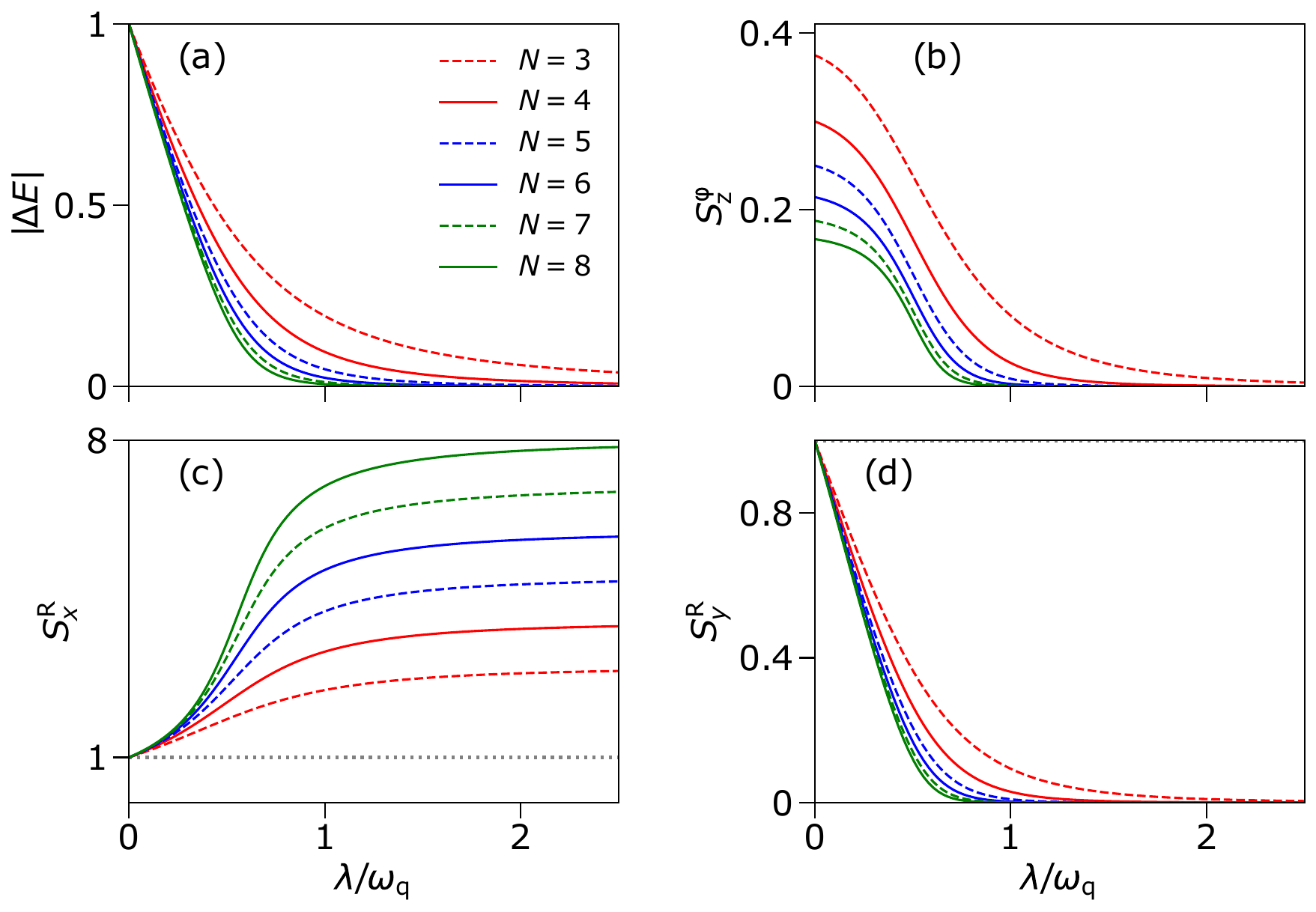}
  \caption{Quantum Ising model: (a) Energy difference between the excited and ground states as a function of the qubit-qubit coupling strength $\lambda$ in unit of the qubit frequency $\omega_{\rm q}$. (b-d) Global susceptibility for pure dephasing $S_{z}^\varphi$ (b) and relaxation, $S^{\rm R}_{x}$ (c) and $S^{\rm R}_{y}$ (d), in $\hat\sigma_{\rm x}$ and $\hat\sigma_{\rm y}$, respectively.}
  \label{FigXX}
\end{figure*}

Significant progress has been made in the realization of quantum computers, primarily utilizing superconducting circuits and ion traps \cite{buluta2011natural,krantz2019quantum,kockum2019,kjaergaard2020,Monroe:2021aa,kwon2021gate,cheng2023noisy,putterman2025hardware}. Other promising platforms, such as trapped Rydberg atoms, spin qubits in semiconductors, and photonic systems, are also actively being explored \cite{bluvstein2024logical,ten2025observation,larsen2025integrated,bordin2025enhanced}. Despite these efforts, none of these platforms are yet capable of performing practical calculations. Among others, the main issue hindering their development is the inevitable interaction with the environment, which leads to decoherence \cite{siddiqi2021engineering}.
Quantum error correction is the way to achieve fault-tolerant quantum computation, but it is fundamental to have more reliable physical qubits that are protected against external noise.

Using superconducting circuits, physical qubits can be protected at the hardware level \cite{gyenis2021moving}. For instance, Cooper pair box, when is capacitively shunted, is protected against charge noise in the transmon regime, drastically increasing its coherence time.
There are two opposite hardware-level strategies to protect qubits: one makes use of a few large elements in specific configurations, such as bifluxon and $0-\pi$ qubit \cite{kalashnikov2020bifluxon,gyenis2021experimental,you2021pokemon};  the other uses a large number of elements in a specific arrangement \cite{douccot2012physical}. Both strategies are challenging due to their stringent energy and configuration requirements.
The last approach results in symmetry-protected ground states, also known as decoherence-free subspaces \cite{lidar1998decoherence,you2010quantum}. A notable example is the 1D Kitaev model of spinless fermions \cite{kitaev2003fault}, which exhibits a double degenerate ground state. This state possess topological nature and is characterized by a pair of localized Majorana edge modes \cite{you2014encoding}. 
The Kitaev model can be mapped to a transverse Ising spin model using the Jordan-Wigner transformation, but, after the transformation, it becomes susceptible to local symmetry-breaking noise.

In general, these protected systems require a high degree of symmetry, such that their states are not sensitive to any slow perturbation (topological protection) \cite{douccot2012physical}. Drawbacks of these protected states are that they are degenerate and also insensible to external manipulation. Preparing and reading out quantum states, as well as performing single- and two-qubit gates for quantum computation, requires a perturbation that resolves the degeneracy and, consequentially, the protection. However, topological quantum computation offers a theoretical framework for performing calculations involving protected degenerate quantum states, but its practical implementation remains a distant goal.

By using superconducting circuits, it has been achieved the ultrastrong coupling regime between artificial atoms and electromagnetic modes \cite{Niemczyk:2010gv,Yoshihara:2017bia,forn2017ultrastrong,kockum2019g,forn2019,qin2024quantum}. Since this regime is implemented on the same platform as some quantum computers, it can be exploited for quantum computing. For instance, \cite{nataf2011protected} propose an array of fluxonium qubits ultrastrongly coupled with an $LC$ resonator to enhance their coherence time; \cite{stassi2018long} proposed to couple longitudinally a flux qubit to a resonator in the ultrastrong coupling regime to increase its relaxation time.

In this article, we propose to couple three or more physical qubits in the ultrastrong coupling regime to form a logical qubit. We demonstrate that, in contrast to the Ising model, alternating interactions in the X and Y directions \emph{suppress pure dephasing and relaxation rates} compared to those of a single physical qubit. This suppression results in \emph{longer coherence times} as either the coupling strength or the number of qubits increases. Furthermore, our model is shown to be more robust against symmetry-breaking noise, a key vulnerability of the Ising model in the ultrastrong coupling regime.

Our results, supported by numerical simulations performed using the QuTiP library \cite{johansson2012qutip}, demonstrate the feasibility of both single- and two-qubit gates with high fidelity. As a possible circuit realization, we propose a design with three flux qubits that realizes an ultrastrong XX interaction between the first two qubits via a Josephson junction and a YY interaction between the second and third qubits through a shared capacitor.

\section{Theoretical model and results}

The coherence time of a quantum system is affected by the interaction of the system with the environment giving rise to two processes: pure dephasing and relaxation. The first is determined by energy fluctuations of the system and affects the relative phases. The relaxation is driven by environmental fluctuations that induce atomic transitions, leading to the loss of energy.

%TODO Need to justify these formula rates.
Considering two eigenstates $\{\vert m\rangle, \vert n\rangle\}$, of a quantum system, the interaction of the $i$th element of the system with its environment (which induces pure dephasing and relaxation) is described by a set of operators $\hat S_{k,i}$ \cite{stassi2018long}, the specific form of which depends on the quantum system under consideration.
Defining $S_{k,i}(m,n)=\langle m \vert \hat S_{k,i} \vert n \rangle$, for each operator $\hat S_{k,i}$, the pure dephasing rate is proportional to 
\begin{equation}
  S^{\rm\varphi}_{k,i}\left(m,n\right)=\frac{1}{4} \left| S_{k,i}(m,m) - S_{k,i}(n,n) \right|^2\,,
\end{equation}
while the relaxation rate is proportional to 
\begin{equation}
  S_{k,i}^{\rm R}\left(m,n\right)=\left| S_{k,i}(m,n) \right|^2\,.
\end{equation}
We refer to $S^{\rm \varphi}_{k,i}\left(m,n\right)$ and $S^{\rm R}_{k,i}\left(m,n\right)$ as \emph{pure dephasing} and \emph{relaxation susceptibilities}, respectively, of the $i$th element of the system to noise in the $\hat{S}_{k,i}$ {\it channel} between the eigenstates $\{\vert m\rangle, \vert n\rangle\}$.

Now, considering the $i$th qubit in a chain of two-level systems, the environmental interactions are described by the set of the Pauli operators: 
\begin{equation}
  \hat S_{k,i}\in\{\hat{\sigma}^{(i)}_x, \hat{\sigma}^{(i)}_y,\hat{\sigma}_z^{(i)}\}\,,
  \label{Pauli}
\end{equation}
with $k=x,y,z$.
If the qubits interact with each other, each Pauli operator in Eq.~(\ref{Pauli}) represents a {\it channel} that couples the {\it entire} system to the environment. 
To simplify our analysis, we define a global pure dephasing and relaxation susceptibilities for each direction $k$ as
\begin{align}
S_{k}^{\rm \varphi} &= \sum_{i} S^{\rm \varphi}_{k,i}
\end{align}
and
\begin{align}
S_k^{\rm R} &= \sum_{i} S^{R}_{k,i}\,,
\end{align}
respectively. These choices are justified under the assumption of a uniform noise spectral density for each direction $k$, and identical parameters for all qubits.
%ToDo mention later in the paper that in the edge susceptivity is different

Ideally, a system is fully protected from environmental noises if $S_{k}^{\rm \varphi}=0$ and $S_k^{\rm R}=0$ for all $k$. Symmetries within the system can protect against specific noise channels. For instance, a two-level system formed by a symmetric double-well potential energy (exhibiting parity symmetry) is insensible to noise in the $\hat\sigma_x$ and $\hat\sigma_y$ channels for pure dephasing, and in the $\hat\sigma_z$ channel for relaxation, such as in the optimal point of a flux qubit \cite{yoshihara2006decoherence}.
In this case, it is straightforward to prove that $S_{z}^{\rm R}=S_x^{\rm \varphi}=S_y^{\rm \varphi}=0$, while the susceptibility to other channels is maximal, $S_{z}^{\rm \varphi}=S_x^{\rm R}=S_y^{\rm R}=1$.

\subsection{The quantum Ising model}

\begin{figure}[hb]
  \includegraphics[scale=0.5]{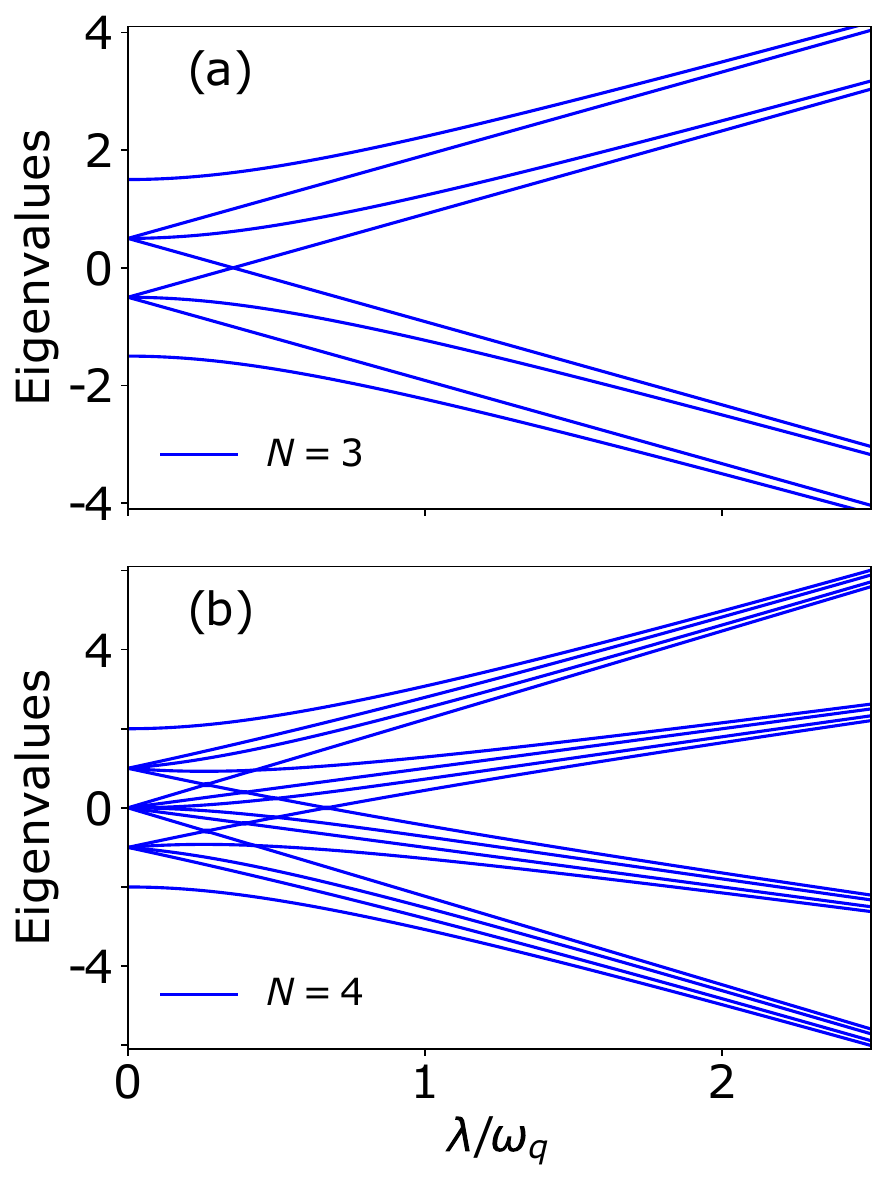}
  \caption{Eigenvalues of the Hamiltonian $\hat H$ as a function of the coupling strength $\lambda$.}\label{eigenVXY}
\end{figure}

We start our analysis by examining an open chain of $N$ artificial atoms (two-level systems) that interact solely along the X direction \cite{you2014encoding,weisbrich2019decoherence,mi2022noise}. This setup corresponds to the Ising model, with its Hamiltonian given by $H_{\rm I}=H_0+H_{\rm XX}$, with
\begin{equation}\label{Hfree}
  \hat H_{0} = -\frac{1}{2} \sum_{i=1}^{N} \omega_{{\rm q}i}\, \hat\sigma_z^{(i)}\,
\end{equation}
and
 \begin{equation}\label{Ising}
  \hat H_{\rm XX} =  - \lambda\sum_{\langle i,j\rangle}\hat\sigma_x^{(i)} \hat\sigma_x^{(j)}\,.
\end{equation} 
Equation~(\ref{Hfree}) represents the atomic energies and, as discussed above, it exhibits parity symmetry. Equation~(\ref{Ising}) is the interaction between nearest neighbors. For simplicity, we consider that all the qubits have same frequency $\omega_{{\rm q}i}=\omega_{\rm q}=1$ and coupling constant $\lambda$.
 
The interaction term in Eq.\,(\ref{Ising}) preserves the parity symmetry of the two-level atoms. As a result, the overall system remains unaffected by noise in the $\hat\sigma_x^{(i)}$ and $\hat\sigma_y^{(i)}$ channels under pure dephasing, and in the $\hat\sigma_z^{(i)}$ channel under relaxation, for all qubits ($i=1,...,N$). When the system enters the ultrastrong coupling regime, the parity symmetry of the interaction term leads to a quasi-double-degenerate ground state [see Fig.~\ref{FigXX}(a)]. These two lowest quasi-degenerate states,
\begin{equation}
	\begin{aligned}
	 \vert \tilde0\rangle &\approx\vert \uparrow	\uparrow..\uparrow\rangle-\vert \downarrow	\downarrow..\downarrow\rangle, \\
	 \vert \tilde1\rangle &\approx\vert \uparrow	\uparrow..\uparrow\rangle+\vert \downarrow	\downarrow..\downarrow\rangle,
   \end{aligned}\label{EqXX}
\end{equation}
are robust against pure dephasing, but are highly susceptible to relaxation. Here $\mid \uparrow\rangle$ and $\mid \downarrow\rangle$ are the eigenstates of the $\hat\sigma_x$ operator. Additionally, small perturbations in the $\hat\sigma_x$ channel break the parity symmetry and the protection.

\begin{figure*}[ht]
  \centering
  \includegraphics[scale=0.42]{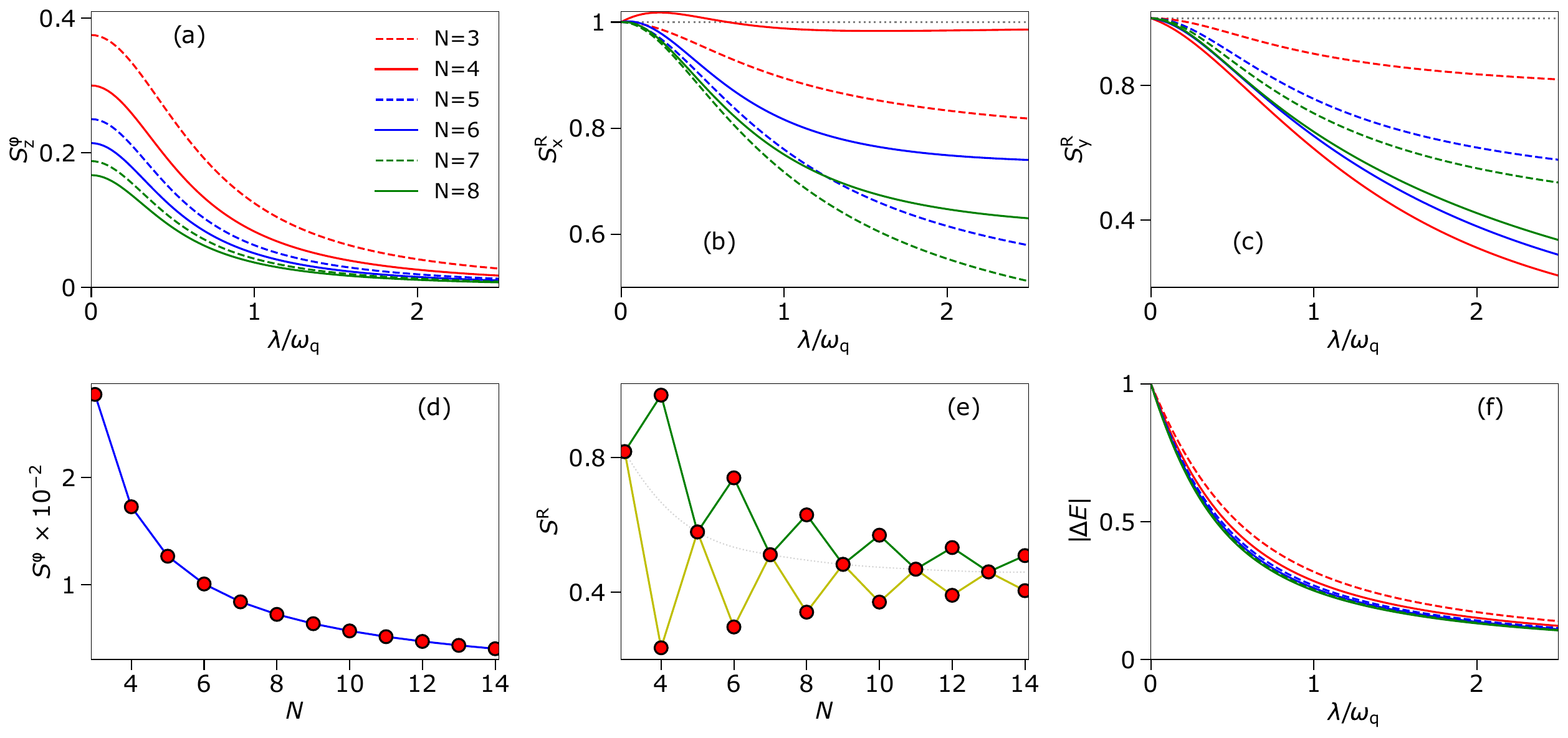}
  \caption{Quantum XY model: (a-c) Global susceptibility (for pure dephasing $S_{z}^\varphi$ and relaxation, $S_{x}^{\rm R}$ and $S_{y}^{\rm R}$, in $\hat\sigma_{\rm x}$ and $\hat\sigma_{\rm y}$) as a function of the coupling strength $\lambda$. (d,e) Global susceptibilities $S^{\varphi}$ and $S^{\rm R}$ as a function of the number $N$ of qubits. (f) Energy difference $\Delta E$ between the excited and ground states.}\label{FigXYN}
\end{figure*}

Figures \ref{FigXX}(b-d) illustrate the susceptibility to pure dephasing and relaxation as a function of the coupling $\lambda$ for a chain of $N$ physical qubits. The numerically calculated susceptibility can be directly compared to that of a single two-level system (qubit), which corresponds to the case where $\lambda/\omega_q=0$ ($S^{\varphi}_z=1$, $S^{R}_x=1$, $S^{R}_y=1$). For large coupling, the susceptibility to noise for pure dephasing $S^{\varphi}_z$ rapidly approaches zero [see Fig.~\ref{FigXX}(b)]. Similarly, the susceptibility for relaxation $S^{R}_y$ in the $\hat\sigma_y$ channel also vanishes quickly [see Fig.~\ref{FigXX}(d)]. In contrast, the susceptibility for relaxation in the $\hat\sigma_x$ channels increases and converges to $N$ [see Fig.~\ref{FigXX}(c)]. This behavior arises because, for large coupling strength ($\lambda \gg \omega_{\rm q}$), the system symmetries do not protect the $\hat\sigma_x$ channels. Moreover, as the eigenstates become maximally entangled [see Eq.~(\ref{EqXX})], all qubits contribute equally to the total susceptibility in this channel.

\subsection{The quantum XY Model}

The system that we propose consists of an open chain of $N$ artificial atoms (two-level systems) with interactions alternating between the X and Y directions \cite{brzezicki2007quantum,sen2010spin,xu2023creating}, as shown in Fig.~\ref{Fig1}.  The Hamiltonian reads $\hat H=\hat H_0+\hat H_{\rm XY}$, with
%$$ H = -\frac{1}{2} \sum_{i=1}^{N} \epsilon_i\, \sigma_z^{(i)} - g \sum_{i=1}^{\left\lfloor \frac{N}{2} \right\rfloor} \sigma_x^{(2i-1)} \sigma_x^{(2i)} +  \sigma_y^{(2i)} \sigma_y^{(2i+1)}, $$
\begin{equation}\label{Hamil}
  \hat H_{\rm XY} = -\lambda\left(\sum_{i=1\,\left(\text{odd}\right)}^{N-1} \hat\sigma_x^{(i)} \hat\sigma_x^{(i+1)}  + \sum_{i=2\, \left(\text{even}\right )}^{N-1} \hat\sigma_y^{(i)} \hat\sigma_y^{(i+1)}\right)\,.
\end{equation}

If the number $N$ of qubits is even, the eigenstates of Eq.~(\ref{Hamil}) form $2^{N/2}$ manifolds. Each manifold consists of $2^{N/2}$ eigenstates  \cite{xu2023creating} that converge to degeneracy when $\lambda/\omega\rightarrow\infty$, due to the presence of non local $N/2$ SU(2) symmetries [see Fig.~\ref{eigenVXY}(a)]. 

Conversely, if the number of qubits is odd, the eigenstates of Eq.~(\ref{Hamil}) form $2^{(N-1)/2}$ manifolds, each containing $2^{(N+1)/2}$ eigenstates that become degenerate when $\lambda/\omega\rightarrow\infty$ [see Fig.~\ref{eigenVXY}(b)]. 

We define the two lowest eigenstates of the total Hamiltonian in Eq.~(\ref{Hamil}) as {\it logical} qubit states $\{\vert 0\rangle, \vert 1\rangle\}$, formed by $N$ strongly interacting physical qubits. Figure \ref{FigXYN}(f) displays the relative energy between the states $\vert 0\rangle$ and $\vert 1\rangle$, that converge to zero when $\lambda/\omega\rightarrow\infty$.

As in the Ising model, the interaction terms preserve the parity symmetry of the two-level atoms. Consequently, the overall system remains unaffected by noise in the $\hat{\sigma}_x^{(i)}$ and $\hat{\sigma}_y^{(i)}$ channels for pure dephasing, as well as in the $\hat{\sigma}_z^{(i)}$ channel for relaxation. However, unlike the Ising model, increasing the coupling $\lambda$ in the XY model, enhances the protection of the two lowest-energy states $\{\vert 0\rangle, \vert 1\rangle\}$ against both pure dephasing and relaxation. This occurs because a higher system symmetry now provides protection against pure dephasing and relaxation in all channels, including $\hat{\sigma}_x$.

Figures~\ref{FigXYN}(a-c) illustrate the pure dephasing and relaxation susceptibility for logical qubits as a function of the coupling $\lambda$ for different numbers of physical qubits $N$. We observe \emph{a reduction in both pure dephasing and relaxation susceptibility} as $\lambda$ or the number of physical qubits $N$ increases for all the channels, including the $\hat\sigma_x^{(i)}$ channels. In comparison to the Ising model, the pure dephasing susceptibility and reduction in energy difference converge to zero in the XY model at a slower rate as a function of $\lambda$ [see Fig.~\ref{FigXYN}(a,f)]. 

\begin{figure*}[ht]
\centering
  \includegraphics[scale=0.5]{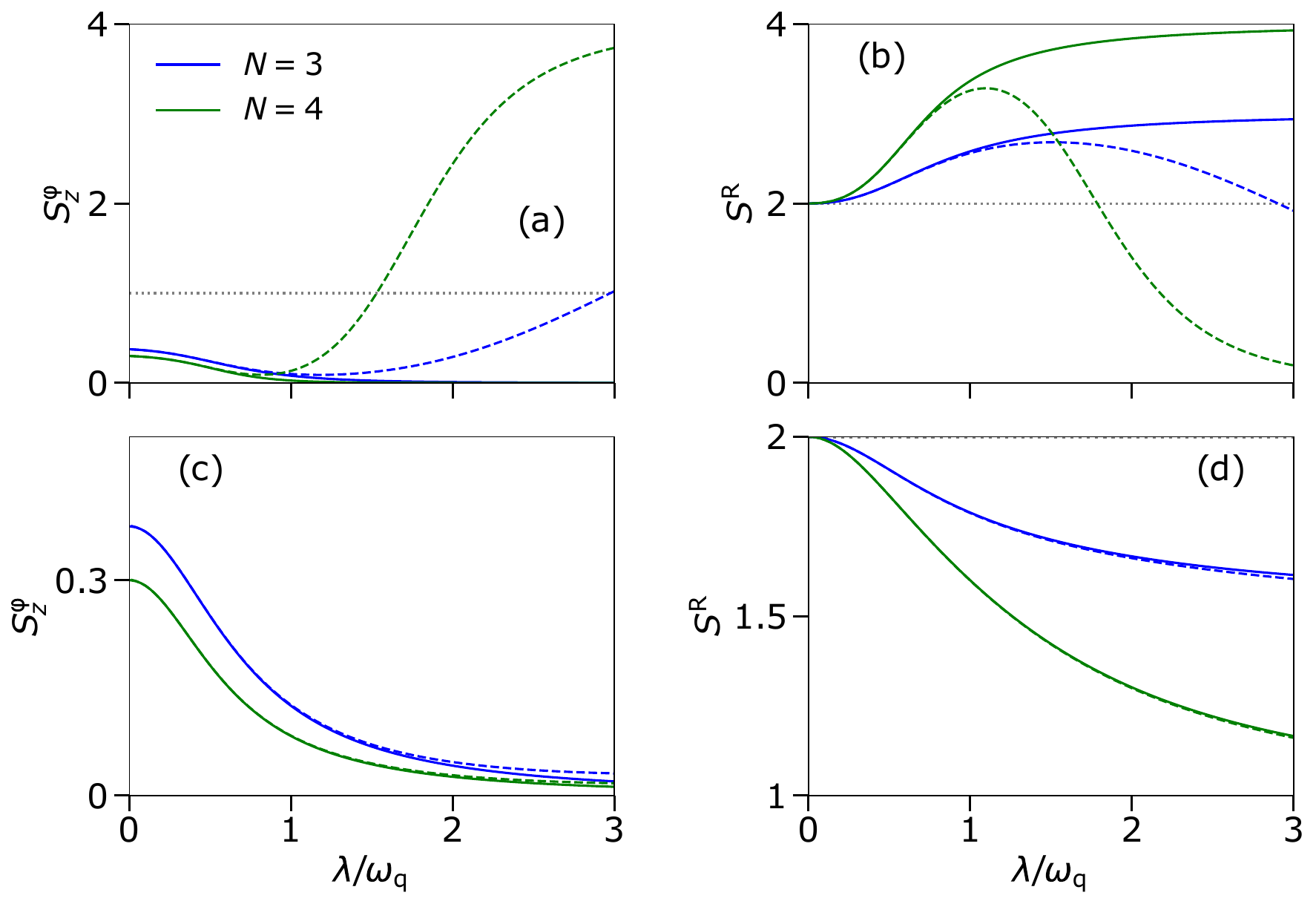}
  \caption{Impact of small perturbations $\hat\sigma_x$ on qubit coherence. (a,b) Quantum Ising model: Susceptibility to pure dephasing $S_{z}^\varphi$ (a) and relaxation $S^{\rm R}$ (b). (c,d) Quantum XY model: Susceptibility $S_{z}^\varphi$ to pure dephasing (c) and $S^{\rm R}$ relaxation (d). Solid curves represent the case without perturbation; dashed curves, with perturbation.}\label{FigXXY_Sym}
\end{figure*}

Figure \ref{FigXYN}(d,e) shows the total susceptibility to noise for pure dephasing and relaxation as a function of the number $N$ of physical qubits, with a fixed coupling strength $\lambda=2.5\,\omega_{\rm q}$. Here, we defined $S^{\rm R}=\sum_k S_k^{\rm R}$ and $S^{\rm \varphi}=\sum_k S_k^{\rm \varphi}$. For $N=9$, the total susceptibility for pure dephasing $S_z^{\varphi}$ is approximately $6\times 10^{-3}$, and for relaxation, it is around $0.48$ for both $\hat S_x^{\rm R}$ and $\hat S_y^{\rm R}$.

Figure \ref{FigXYN}(e) reveals that relaxation suppression differs between the $\hat\sigma_x$ and $\hat\sigma_y$ operators when the number of physical qubits, $N$, is even. This asymmetry arises because, for an even $N$, interchanging $\hat\sigma_x$ and $\hat\sigma_y$ operators changes the interaction Hamiltonian in $\hat H$. Conversely, for an odd $N$, swapping $\hat\sigma_x$ and $\hat\sigma_y$ and taking the mirror image of the chain leaves the Hamiltonian unchanged.
For $N=4$, the system seems not protected in the $\hat\sigma_x$ channel for relaxation, as $S^{\rm R}\approx 1$. However, it is protected, considering that in the Ising model [see Fig.\,\ref{FigXX}(c)], the noise in this channel increases with increasing the coupling strength $\lambda$.

\subsection{Robustness to Symmetry-Breaking Noise}

Figure \ref{FigXXY_Sym} displays the susceptibility to pure dephasing and relaxation as a function of the coupling strength $\lambda/\omega_{\rm q}$. It shows results for both the Ising model [Figs.~\ref{FigXXY_Sym}(a,b)] and the proposed XY model [Figs.~\ref{FigXXY_Sym}(c,d)] for $N=3,4$. Dashed curves indicate the case with a small $\hat\sigma_x$ perturbation, set to $1\%$ of the bare atomic energy, while solid curves represent the unperturbed scenario.

When the coupling is deep in the ultrastrong coupling regime ($\lambda>\omega_q$), the eigenstates of the Ising model, and consequently its susceptibility, are highly sensitive to small fluctuations in the $\hat\sigma_x$ channel. In fact, in the Ising model, the susceptibilities for pure dephasing and relaxation change dramatically for $\lambda>\omega_{\rm q}$ [see Figs.~\ref{FigXXY_Sym}(a,b)], indicating instability due to
local perturbations. On the contrary, the XY model exhibits robustness in the same regime, showing only minor variations in its susceptibilities [see Figs.~\ref{FigXXY_Sym}(c,d)].

\section{Dynamical evolution and gates}
\subsection{Comparing Single Physical Qubit and Logical Qubit Dynamics}
We numerically simulate the evolution of a logical quantum state formed by $N=4$ qubits interacting in the ultrastrong coupling regime, $\lambda=1.3\,\omega_{\rm q}$, in the XY configuration, considering the interaction with the environment. We then compare this with the evolution of a single qubit subjected to the same environment. The evolution is governed by a master equation in the Lindblad form \cite{beaudoin2011dissipation,stassi2018long}, applied after diagonalizing the full Hamiltonian $\hat H$,

\begin{eqnarray}\label{MasterEQ}
  \dot{\hat\rho} &=&	-i\left[\hat {H},\hat\rho\right]+\sum_{k,i}\sum_{m,\,n>m}\Gamma^{(k,i)}_{mn}\mathcal{D}\left[\vert m\rangle\langle n\vert\right]\hat\rho\\ 
  &&+\frac{1}{2}\sum_{k,i}\gamma^{(k,i)}_{\rm \varphi}\mathcal D\left[\sum_mS_{k,i}(m,m) \vert m\rangle\langle m\vert\right]\hat\rho\,,\nonumber
\end{eqnarray}
where $\mathcal{D}[\hat O]\hat\rho=(2\hat O\hat\rho\,\hat O^\dagger-\hat O^\dagger\hat O\hat\rho-\hat\rho\,\hat O^\dagger\hat O)/2$ is the Lindblad superoperator and $\hat\rho$ is the density matrix. The sum over $k$ and $i$ accounts all decoherence channels $\hat{S}_{k,i}$. The transition rates from eigenstate $\vert n\rangle$ to eigenstate $\vert m\rangle$ are given by $\Gamma^{(k,i)}_{mn}=\gamma^{(k,i)}(\omega_{mn})S_{k,i}^{\rm R}\left(m,n\right)$, where $\omega_{mn}=\omega_{m}-\omega_{n}$. The functions $\gamma^{(k,i)}(\omega_{mn})$ and $\gamma^{(k,i)}_{\rm \varphi}$ are proportional to the noise spectral density, and correspond to the relaxation and pure dephasing rates, respectively, in the absence of ultrastrong coupling. 
The second term on the right-hand side of Eq.~(\ref{MasterEQ}) describes relaxation, while the third term accounts for pure dephasing.  Expanding the last term in the master equation (\ref{MasterEQ}) shows that the pure dephasing rate for each channel in the ultrastrong coupling regime is given by $\tilde\gamma^{(k,i)}_{\rm \varphi}=\gamma^{(k,i)}_{\rm \varphi} S^{\rm\varphi}_{k,i}\left(m,n\right)$ \cite{mercurio2023pure}.

\begin{figure}[t]
  \centering
  \includegraphics[scale=0.5]{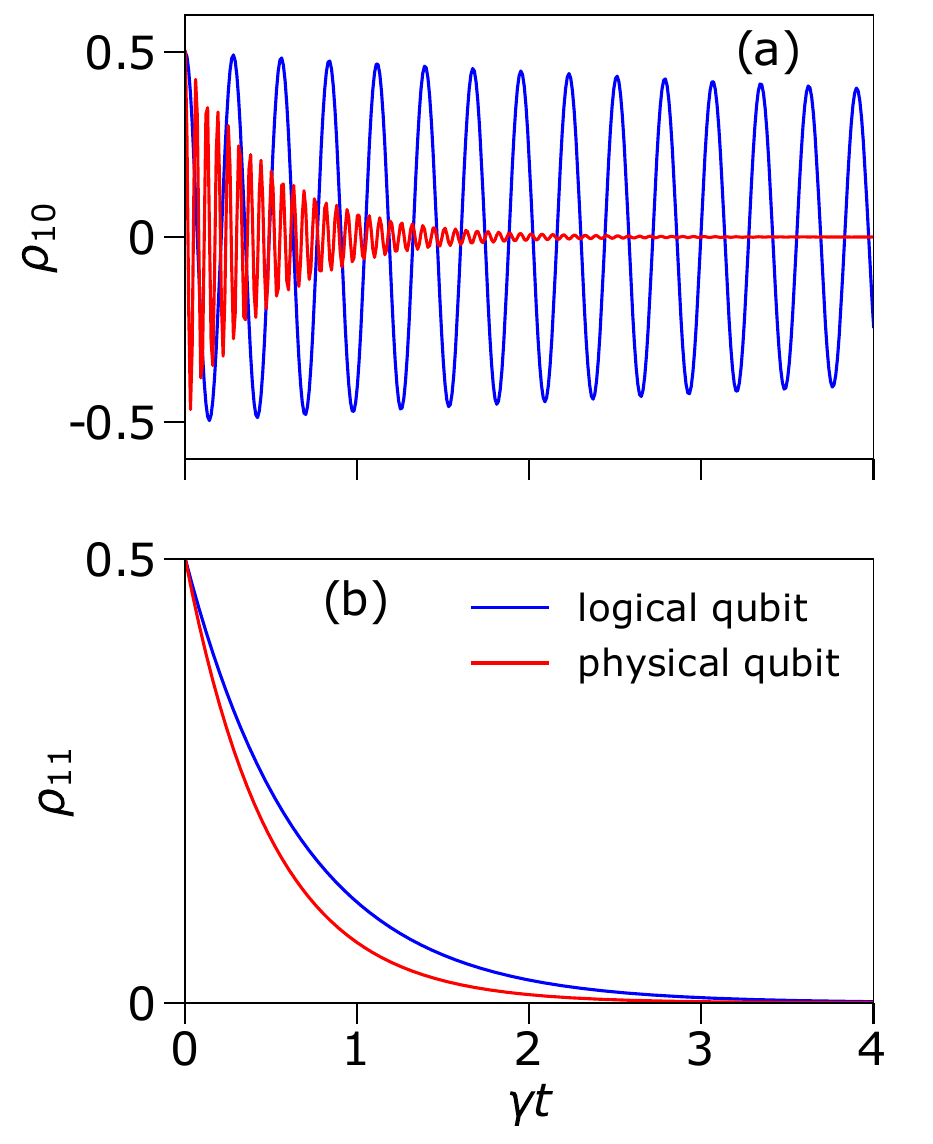}
  \caption{Pure dephasing $\rho_{10}$ (a) and relaxation processes $\rho_{11}$ (b) for a single non-interacting qubit (red curves) and the logical qubit (blue curves) formed by four physical qubits ultrastrongly coupled in the XY configuration. Here $\gamma=\gamma_{\varphi}=10^{-2}$.}
  \label{FigXY_PR}
\end{figure}

Figures \ref{FigXY_PR}(a) and \ref{FigXY_PR}(b) display  $\rho_{11}=\langle 1\vert\hat\rho \vert 1 \rangle$ and $\rho_{10}=\langle 1\vert\hat\rho\vert 0 \rangle$ as a function of time, showing relaxation and pure dephasing processes, respectively, when the initial system state is $\vert \psi\rangle=(\vert 0\rangle+\vert 1\rangle)/\sqrt{2}$. The red curve represents a non-interacting qubit, while the blue curve corresponds to the logical qubit formed by the ultrastrongly coupled system. In Fig.~\ref{FigXY_PR}(a), the relaxation process is excluded to highlight pure dephasing. Conversely, in Fig.~\ref{FigXY_PR}(b), the pure dephasing process is excluded to highlight relaxation. The relaxation and pure dephasing rates for all the physical qubits $i$ and for all the directions $k$, in the absence of ultrastrong coupling, are set to the same value: $\gamma=\gamma^{(k,i)}=10^{-2}$ and $\gamma_{\varphi}=\gamma^{(k,i)}_{\rm \varphi}=10^{-2}$. By comparing the red and blue curves, one can notice the advantage of utilizing strongly interacting physical qubits over a single physical qubit. Specifically, the gain obtained for pure dephasing is $G_\varphi= (S_z^{\varphi})^{-1}=18.05$, while the gain obtained for relaxation is $G_{\rm R}=2(S_x^{\rm R}+S_y^{\rm R})^{-1}=1.34$. We define the gain as $G_\varphi=\gamma_{\rm \varphi}/\tilde\gamma_{\varphi}$ and $G_{\rm R}=\gamma/\tilde\gamma$, where $\tilde\gamma_{\varphi}$ ($\tilde\gamma$) represents the pure dephasing (relaxation) rate of the logical qubit.
 
 \begin{figure}[b]
  \includegraphics[scale=0.5]{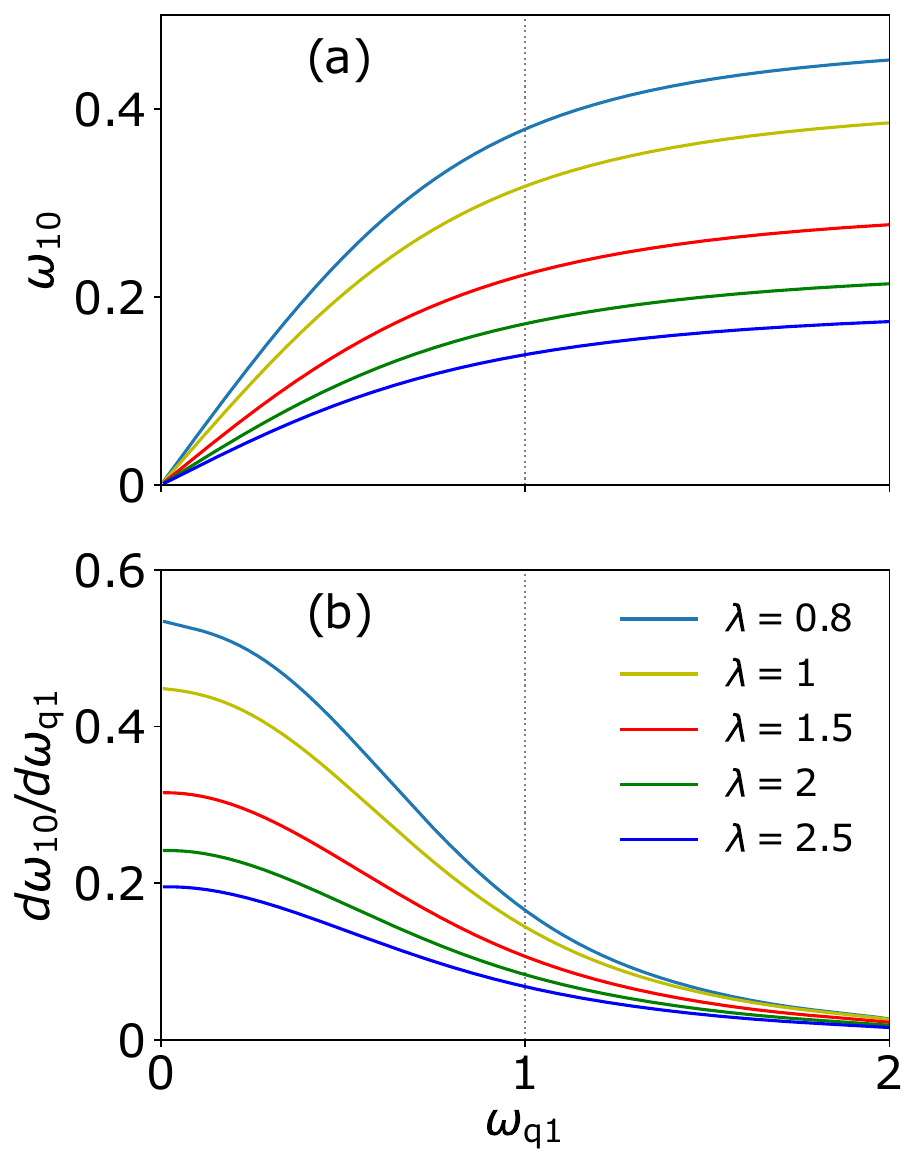}
  \caption{Z gate. (a) Logical qubit frequency $\omega_{10}=\omega_{1}-\omega_{0}$ as a function of single qubit frequency $\omega_{\rm q1}$ for different values of $\lambda$. (b) Sensitivity of the logical qubit frequency to the variation of $\omega_{\rm q1}$.}\label{sgmz}
\end{figure}
 
\subsection{Single- and Two-Qubit Gates}
In this section, we demonstrate that single- and two-qubit gates can be performed with a high fidelity using the introduced logical qubit.
 
The single-qubit X and Y gates can be implemented on the logical qubit by applying a $\pi$-pulse to one of the physical qubits. The Z gate can be realized by modulating the frequency of one of the physical qubits.
Two-qubit gates are implemented by inductively coupling two physical qubits, each belonging to distinct logical qubits. This coupling provides an XX interaction between two logical qubits, resulting in a $\sqrt{\rm iSWAP}$ gate.

Numerical simulations were performed on a system of $N=3$ physical qubits in the ultrastrong coupling regime, with coupling strength ranging from $\lambda=0.5\,\omega_{\rm q}$ to $2.5\,\omega_{\rm q}$, and performed over a large ensemble of random initial states. Simulations demonstrate an \emph{average gate fidelity of $99.99\%$ for both single- and two-qubit gates}. To highlight the intrinsic performance of the gates within our logical qubit, decoherence effects were not included in the simulation.

Figure \ref{sgmz}(a) shows the logical qubit frequency, $\omega_{10}=\omega_{1}-\omega_{0}$, as a function of the modulated physical qubit frequency $\omega_{\rm q1}$. This demonstrates that varying $\omega_{\rm q1}$ around a reference value (here $\omega_{\rm q1}=1$) induces a change in $\omega_{10}$, effectively realizing a Z gate. The derivative, shown in Fig.\,\ref{sgmz}(b), reveals an asymmetric sensitivity of $\omega_{10}$ to variations in $\omega_{\rm q1}$, with decreases in $\omega_{\rm q1}$ having a stronger effect than increases.

\section{Quantum Circuit implementation}
%ToDo How capacitive coupling can reach the USC regime?
Before providing any practical implementation, it is important to experimentally validate the proposed scheme. The ultrastrong coupling regime between a qubit and a resonator has already been achieved using superconducting circuits \cite{Niemczyk:2010gv,tomonaga2025spectral, wang2024strong}. In this platform, such large coupling strengths can be realized when the qubit and a resonator share a common line interrupted by a Josephson junction. 

An ultrastrong XX interaction is a straightforward extension of this scheme to two flux qubits \cite{you2007low,yan2016flux}.
A YY interaction can be implemented by coupling two flux qubits via a shared capacitor. While there have been recent theoretical investigations \cite{hita2021three,hita2022ultrastrong}, the ultrastrong YY coupling —either between qubits or between a qubit and a resonator— remains experimentally unexplored. 
%The main challenge lies in designing a quantum circuit with parameters that predominantly yields a YY interaction while suppressing parasitic XX and ZZ couplings. 
Demonstrating this form of coupling would not only allow to prove the validity of our proposal but also open new pathways for quantum simulations of spin models and the implementation of nontrivial entangling gates.

\begin{figure}[hbt]
  \includegraphics[scale=0.69]{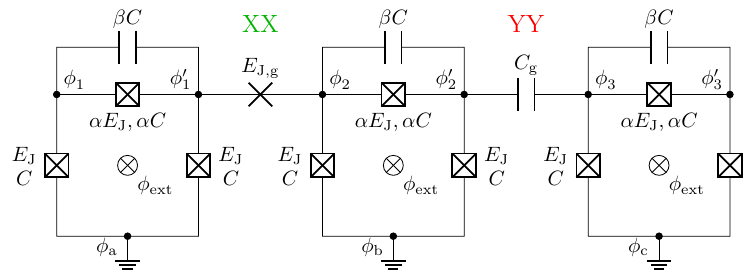}
  \caption{Diagram of three interacting flux qubits. The first two qubits are coupled via a Josephson junction of energy $E_{\rm J,g}$, while the second and third qubits are coupled via a capacitor $C_ {\rm g}$.}\label{circuit}
\end{figure}
Figure \ref{circuit} shows a possible circuit implementation to verify the validity of this proposal. The circuit is composed by three flux qubits: the first two are inductively coupled, while the second and the third are capacitively coupled. A Josephson junction is inserted in the common line between the first and  second artificial qubits, to increase the coupling to the ultrastrong coupling regime. With these circuit elements, the first and second qubits realize an XX interaction, while the second and the third YY interaction. A complete derivation of the Hamiltonian circuit is provided in Appendix \ref{Circhamil}.
\\
\section{discussion}

The results presented in this work demonstrate that three or more qubits interacting in the ultrastrong coupling regime in an alternating XX and YY interactions outperform a single qubit in preserving coherence.

Using the Lindblad master equation, our numerical simulations reveal a significant suppression of both pure dephasing and relaxation rates for the logical qubit, leading to an increase in coherence time. This protection originates from the symmetries of the alternating XY interaction Hamiltonian in the ultrastrong coupling regime.

We further show that high-fidelity single- and two-qubit gates can be implemented on the logical qubit, enabling universal quantum computation within this framework. 

Experimental validation of the predicted suppression of noise susceptibility is a critical next step. This can be directly tested using flux qubits, which are well-suited for realizing ultrastrong coupling. While the generation of ultrastrong XX coupling is well-established in flux qubit architectures, the implementation of ultrastrong YY coupling remains an experimental challenge. Nevertheless, recent theoretical advancements in circuit design offer promising avenues for achieving this interaction with minimized parasitic couplings.
 
Beyond flux qubits, it is of significant importance to investigate the behavior of transmons in the ultrastrong coupling regime. Given their leading coherence times among superconducting artificial atoms, extending this noise protection scheme to transmon-based systems could pave the way for even longer-lived logical qubits and accelerated progress towards fault-tolerant quantum computing. 

\appendix
\section{Eigenstates of Hamiltonian $\hat H$}

Here, we derive the eigenstates for \( N=3 \) and 4 in the strong-coupling limit \(\lambda \to \infty\).  

\subsection{$N=3$ Atoms}  
For \( N=3 \) atoms, the four lowest eigenstates of the Hamiltonian $\hat H$, that become degenerate as \( \lambda \to \infty \), are:

\begin{widetext}
\begin{equation}
\begin{array}{rl}
    \ket{\Tilde{0}} &= \frac{1}{2} e^{i\pi/2} \big[ -\ket{\downarrow \; \swarrow\; \leftarrow} + \ket{\uparrow \; \nearrow \; \leftarrow} 
    - \ket{\uparrow \; \nwarrow \; \rightarrow} + \ket{\downarrow \; \searrow \; \rightarrow} \big], \\[8pt]
    
    \ket{\Tilde{1}} &= \frac{1}{2} \big[ e^{-i\pi/4} (-\ket{\downarrow \; \swarrow \; \leftarrow} + \ket{\uparrow \; \nwarrow \; \rightarrow}) 
    + e^{i\pi/4} (\ket{\uparrow \; \nearrow \; \leftarrow} - \ket{\downarrow \; \searrow \; \rightarrow}) \big], \\[8pt]
    
    \ket{\Tilde{2}} &= \frac{1}{2} \big[ e^{i\pi/4} (\ket{\downarrow \; \swarrow \; \leftarrow} - \ket{\uparrow \; \nwarrow \; \rightarrow}) 
    + e^{-i\pi/4} (-\ket{\uparrow \; \nearrow \; \leftarrow} + \ket{\downarrow \; \searrow \; \rightarrow}) \big], \\[8pt]
    
    \ket{\Tilde{3}} &= \frac{1}{2} \big[ \ket{\downarrow \; \swarrow\; \leftarrow} + \ket{\uparrow \; \nearrow \; \leftarrow} 
    + \ket{\uparrow \; \nwarrow \; \rightarrow} + \ket{\downarrow \; \searrow \; \rightarrow} \big].
\end{array}
\end{equation}
\end{widetext}

where \( \{\ket{\uparrow},\ket{\downarrow}\} \) and \( \{\ket{\rightarrow},\ket{\leftarrow}\} \) are the eigenstates of \( \hat{\sigma}_x \) and \( \hat{\sigma}_y \), respectively. The states \( \ket{\nearrow}, \ket{\nwarrow}, \ket{\swarrow}, \) and \( \ket{\searrow} \) are the eigenstates of $m\hat\sigma_x+n\hat\sigma_y$ corresponding to the lowest eigenvalues for all combinations of $(m,n)=(\pm 1)$:

\begin{equation}
\begin{array}{ll}
    \ket{\nearrow} = \frac{1}{\sqrt{2}} (\ket{0} + e^{i\pi/4} \ket{1}), \quad &
    \ket{\swarrow} = \frac{1}{\sqrt{2}} (\ket{0} - e^{-i\pi/4} \ket{1}), \\[8pt]
    \ket{\nwarrow} = \frac{1}{\sqrt{2}} (\ket{0} + e^{-i\pi/4} \ket{1}), \quad &
    \ket{\searrow} = \frac{1}{\sqrt{2}} (\ket{0} - e^{i\pi/4} \ket{1}).
\end{array}
\end{equation}

\subsection{$N=4$ Atoms}

For \( N=4 \) atoms, the four lowest eigenstates of $\hat H$, that become degenerate as \( \lambda \to \infty \), are:

\begin{equation}
\begin{array}{rcl}
\ket{\tilde 0} &=& a_{\rm p} \ket{\phi_1} + a_{\rm m} \ket{\phi_2} + a_{\rm m} \ket{\phi_3} - a_{\rm p} \ket{\phi_4}, \\
\ket{\tilde 1} &=& \frac{1}{\sqrt{2}}\left(\ket{\phi_1} + \ket{\phi_4}\right), \\
\ket{\tilde 2} &=& \frac{1}{\sqrt{2}}\left(\ket{\phi_2} + \ket{\phi_3}\right), \\
\ket{\tilde 3} &=& a_{\rm m} \ket{\phi_1} - a_{\rm p} \ket{\phi_2} - a_{\rm p} \ket{\phi_3} - a_{\rm m} \ket{\phi_4}.
\end{array}
\end{equation}

\noindent where $a_{\rm p,m} =\frac{1}{\sqrt{5\pm\sqrt{5}}}$, and

\begin{equation}
\begin{array}{rcl}
\ket{\phi_{1}} &=& \ket{\downarrow}(-c_1\ket{\uparrow\uparrow} + c_2\ket{\downarrow\downarrow})\ket{\downarrow}, \\
\ket{\phi_{2}} &=& \ket{\downarrow}(c_1\ket{\uparrow\downarrow} + c_2\ket{\downarrow\uparrow})\ket{\uparrow}, \\
\ket{\phi_{3}} &=& \ket{\uparrow}(c_1\ket{\downarrow\uparrow} + c_2\ket{\uparrow\downarrow})\ket{\downarrow}, \\
\ket{\phi_{4}} &=& \ket{\uparrow}(c_1\ket{\downarrow\downarrow} - c_2\ket{\uparrow\uparrow})\ket{\uparrow},
\end{array}
\end{equation}
with $c_1 = a_{\rm m}-a_{\rm p}$ and $c_2 = a_{\rm m}+a_{\rm p}$.

\section{Derivation of the Circuit Hamiltonian}\label{Circhamil}

This appendix provides a comprehensive derivation of the full circuit Hamiltonian in Fig.~\ref{circuit}. We begin in Sec.~\ref{subsection:Derivation_of_the_Lagrangian} by constructing the total Lagrangian of the system, which includes terms for the individual 3JJ flux qubits and their capacitive and inductive inter-qubit couplings. Subsequently, in Sec.~\ref{subsection:Derivation_of_the_Hamiltonian}, we perform a Legendre transformation to obtain the classical Hamiltonian. Finally, in Sec.~\ref{subsection:Quantization_and_two-level_truncation}, we outline the canonical quantization procedure and the two-level truncation necessary to arrive at the effective spin Hamiltonian used in our analysis.
\\
\subsection{Derivation of the Lagrangian}\label{subsection:Derivation_of_the_Lagrangian}

The Lagrangian of a single 3JJ flux qubit (3JJQ), which is the constituent block of Fig.~\ref{circuit}, is given by
\begin{align} \label{eq:L_q_i}
    \mathcal{L}_{{\rm q},i} &= \frac{1}{2} \dot{\boldsymbol{\phi}}_i^T \mathbf{C}_{\rm q} \dot{\boldsymbol{\phi}}_i + E_{\rm J} \cos{\left(\frac{\phi_i}{\Phi_0}\right)} + E_{\rm J} \cos{\left(\frac{\phi_i^\prime}{\Phi_0}\right)} \nonumber \\
    & \quad + \alpha E_{\rm J} \cos{\left(\frac{\phi_i^\prime - \phi_i -\phi_{\rm ext}}{\Phi_0}\right)} \, .
\end{align}
being $\boldsymbol{\phi}_i = \left( \phi_i, \phi_i^\prime \right)^T$ the vector of the flux node variables of the single qubit (with $i=1,2,3$), $\Phi_0 = \hbar /2e$ the reduced flux quantum, $\phi_{\rm ext}$ the constant external flux threading the loop. The capacitance matrix is given by
\begin{equation} \label{eq:C_q}
    \mathbf{C}_q = C \,
    \begin{pmatrix}
        1 + \alpha + \beta & -(\alpha + \beta) \\
        -(\alpha + \beta) & 1 + \alpha + \beta
    \end{pmatrix} \, .
\end{equation}
In the derivation of the Lagrangian of our circuital scheme, the ground nodes of the 3JJQ are chosen to be between the identical Josephson junctions, i.e.  $\phi_{\rm a} = \phi_{\rm b} = \phi_{\rm c} = 0$.

We now consider three identical JJQs coupled through a JJ and a capacitor, as in Fig.~\ref{circuit}. The respective interaction Lagrangians are $\mathcal{L}_{\rm int, JJ} = E_{\rm J, g} \cos{\left(\frac{\phi_3 - \phi_2^\prime}{\Phi_0}\right)}$ and $\mathcal{L}_{\rm int, C} = \frac{C_{\rm g}}{2} \left( \dot{\phi}_2 - \dot{\phi}^\prime_1 \right)^2$, and hence the total Lagrangian is
% Inizio dell'ambiente widetext per la matrice estesa
\begin{widetext}
\begin{align} \label{eq:L_tot_sum}
    \mathcal{L}_{\rm tot} &= \sum_{i=1}^3 \mathcal{L}_{{\rm q},i} + \mathcal{L}_{\rm int, JJ} + \mathcal{L}_{\rm int, C} = \frac{1}{2} \dot{\boldsymbol{\phi}}^T \mathbf{C} \dot{\boldsymbol{\phi}} + \sum_{i=1}^3 E_{\rm J} \left[ \cos{\left(\frac{\phi_i}{\Phi_0}\right)} + \cos{\left(\frac{\phi_i^\prime}{\Phi_0}\right)} \right. \left. + \alpha \cos{\left(\frac{\phi_i^\prime - \phi_i -\phi_{\rm ext}}{\Phi_0}\right)}\right]\\ \nonumber
    &+ E_{\rm J, g} \cos{\left(\frac{\phi_2 - \phi_1^\prime}{\Phi_0}\right)}  + \frac{C_{\rm g}}{2} \left( \dot{\phi}_3 - \dot{\phi}^\prime_2 \right)^2 \, .
\end{align}
\end{widetext}
where $\boldsymbol{\phi} = \left( \boldsymbol{\phi}_1, \boldsymbol{\phi}_2, \boldsymbol{\phi}_3 \right)^T$ is the total flux node vector and $\mathbf{C}$ is the total bare capacitance matrix, which is block diagonal with each block composed by the single qubit $2\times2$ capacitance matrices $\mathbf{C}_{\rm q}$.
The matrix $\tilde{\mathbf{C}}$ is the capacitance matrix renormalized by the inductive coupling $C_{\rm g}.$ An explicit expression of $\tilde{\mathbf{C}}$ is thus given by
% Inizio dell'ambiente widetext per la matrice estesa
\begin{widetext}
\begin{equation} \label{eq:renorm_cap_matrix}
    \tilde{\mathbf{C}} = C \,
    \begin{pmatrix}
        1 + \alpha + \beta & -(\alpha + \beta) & 0 & 0 & 0 & 0 \\
        -(\alpha + \beta) & 1 + \alpha + \beta & 0 & 0 & 0 & 0 \\
        0 & 0 & 1 + \alpha + \beta & -(\alpha + \beta) & 0 & 0 \\ % Correzione della voce [3,3]
        0 & 0 & -(\alpha + \beta) & 1 + \alpha + \beta + \gamma & -\gamma & 0 \\
        0 & 0 & 0 & -\gamma & 1 + \alpha + \beta + \gamma & -(\alpha + \beta) \\
        0 & 0 & 0 & 0 & -(\alpha + \beta) & 1 + \alpha + \beta
    \end{pmatrix} \, ,
\end{equation}
\end{widetext}
where we considered $C_{\rm g} = \gamma C$. We explicitly notice that the upper block (regarding the first flux qubit on the left) is not renormalized by the capacitive interaction.

\subsection{Derivation of the Hamiltonian}\label{subsection:Derivation_of_the_Hamiltonian}

From the total Lagrangian in \cref{eq:L_tot_sum}, we can derive the corresponding Hamiltonian.
We firstly calculate the conjugated momenta, which represent the node charges, that are given by the vector
\begin{equation} \label{eq:Q_def}
    \mathbf{Q} = \frac{\partial\mathcal{L}}{\partial\dot{\boldsymbol{\phi}}} = \tilde{\mathbf{C}} \dot{\boldsymbol{\phi}} \, .
\end{equation}
Therefore, by inverting this equation, we have $\dot{\boldsymbol{\phi}} = \tilde{\mathbf{C}}^{-1} \mathbf{Q}$. The corresponding Hamiltonian $H \left( \mathbf{Q}, \boldsymbol{\phi} \right) = \mathbf{Q}^T \, \dot{\boldsymbol{\phi}} - \mathcal{L} \left( \dot{\boldsymbol{\phi}}, \boldsymbol{\phi} \right)$ is given by
\begin{align} \label{eq:H_tot_explicit} % Eq. 7 nel PDF
    H &= \frac{1}{2} \mathbf{Q}^T \tilde{\mathbf{C}}^{-1} \mathbf{Q} - \sum_{i=1}^3 E_{\rm J} \left[ \cos{\left(\frac{\phi_i}{\Phi_0}\right)} + \cos{\left(\frac{\phi_i^\prime}{\Phi_0}\right)} \right. \nonumber \\
    & \quad \left. + \alpha \cos{\left(\frac{\phi_i^\prime - \phi_i -\phi_{\rm ext}}{\Phi_0}\right)}\right] - E_{\rm J, g} \cos{\left(\frac{\phi_2 - \phi_1^\prime}{\Phi_0}\right)} \, .
\end{align}

We notice that the inverse of the local renormalized capacitance matrix is non-local by construction, as the coupling renormalizes also the resonance frequencies of qubit 2 and qubit 3, while however not affecting the frequency of qubit 1, as it can be easily demonstrated by direct calculation of $\tilde{\mathbf{C}}^{-1}$. Indeed, we can write this inverse as a block matrix of the form
\begin{equation} \label{eq:C_inv_block}
    \tilde{\mathbf{C}}^{-1} =
    \begin{pmatrix}
        \bar{\mathbf{C}}_{{\rm q},1}^{-1} & \mathbf{0} & \mathbf{0} \\
        \mathbf{0} & \bar{\mathbf{C}}_{{\rm q},2}^{-1} & \bar{\mathbf{C}}_{\rm c}^{-1} \\
        \mathbf{0} & (\bar{\mathbf{C}}_{\rm c}^{-1})^T & \bar{\mathbf{C}}_{{\rm q},3}^{-1}
    \end{pmatrix} \, .
\end{equation}
As discussed previously, $\bar{\mathbf{C}}_{{\rm q},1} = \mathbf{C}_{\rm q}$, while the matrices for the other two qubits are affected by the interaction $C_{\rm g}$, the matrix $\bar{\mathbf{C}}_{\rm c}^{-1}$ is obtained by direct calculation of the off-diagonal block of the inverse of the total renormalized capacitance matrix $\tilde{\mathbf{C}}$.

With the introduction of these definitions, we can therefore rewrite the total Hamiltonian as
\begin{equation} \label{eq:H_tot}
    H = \sum_{i=1}^3 H_{{\rm q},i} + H_{\rm int,JJ} + H_{\rm int,C} \, ,
\end{equation}
where
\begin{align} \label{eq:H_qubit}
    H_{{\rm q},i} &= \frac{1}{2} \mathbf{Q}_i^T \bar{\mathbf{C}}_{{\rm q},i}^{-1} \mathbf{Q}_i - E_{\rm J} \left[ \cos{\left(\frac{\phi_i}{\Phi_0}\right)} \right. \nonumber \\
    & \quad \left. + \cos{\left(\frac{\phi_i^\prime}{\Phi_0}\right)} + \alpha \cos{\left(\frac{\phi_i^\prime - \phi_i -\phi_{\rm ext}}{\Phi_0}\right)}\right] \, .
\end{align}
are the single qubit Hamiltonians (notice that the Hamiltonians of the qubits 2 and 3 are renormalized by the capacitive interaction, while qubit 1 is not affected, as pointed out previously), $H_{\rm int,JJ} = - E_{\rm J, g} \cos{\left(\frac{\phi_2 - \phi_1^\prime}{\Phi_0}\right)}$, and $H_{\rm int,C} = - \mathbf{Q}_2 \bar{\mathbf{C}}_{\rm c}^{-1} \mathbf{Q}_3$ are the inductive and capacitive interaction Hamiltonians.

\subsection{Quantization and two-level truncation}\label{subsection:Quantization_and_two-level_truncation} % Cambiato da section a subsection

The canonical quantization procedure for circuit QED promotes the node fluxes and charges in \cref{eq:H_tot} to operators satisfying the canonical commutation relations $\left[ \hat{\phi}_i, \hat{Q}_j \right] = i \hbar \delta_{ij}$.
After performing the two-level truncation in the basis of lowest energy levels of the qubit Hamiltonians $H_{{\rm q},i}$ and working at the symmetry point $\phi_{\rm ext} = \pi \Phi_0$, the total Hamiltonian in \cref{eq:H_tot} becomes
\begin{equation} \label{eq:H_final_qubit}
    \hat{H} = \sum_{i=1}^3 \frac{\omega_{{\rm q}i}}{2} \hat{\sigma}^{(i)}_z \!+ \!\! \sum_{k,l = x,y,z} \!\! \left( g_{{\rm JJ}, kl} \, \hat{\sigma}^{(1)}_k \hat{\sigma}^{(2)}_l \!+ g_{{\rm C}, kl} \, \hat{\sigma}^{(2)}_k \hat{\sigma}^{(3)}_l \right) \, ,
\end{equation}
where $\omega_{{\rm q}i}$ are the (renormalized) single-qubit energy gaps, while $g_{{\rm JJ}, kl}$ and $g_{{\rm C}, kl}$ are the inductive and capacitive coupling constants, respectively, connecting the \textit{k}-th and \textit{l}-th orientations of the adjacent qubits. 

Among the inductive terms, $g_{{\rm JJ}, xx}$ dominates all the other inductive interactions, making the leading contribution proportional to $\hat{\sigma}^{(1)}_x \hat{\sigma}^{(2)}_x$. In contrast, the dominant term in the capacitive interaction (calculated by first-order perturbation theory) is proportional to $\hat{\sigma}^{(2)}_y \hat{\sigma}^{(3)}_y$. Nevertheless, the second-order term (proportional to $\hat{\sigma}^{(2)}_z \hat{\sigma}^{(3)}_z$) can become non-negligible for certain parameters. 

To suppress these contributions, the device can be designed such that the energy gap between the ground and first excited states of the qubits energy spectra is nearly equal to that between higher excited states (particularly the gap between the second and third levels, which is the primary contribution).
Specifically, choosing $\gamma = C_{\rm g}/C \approx 5$ (while fixing $\alpha = 0.65$, $\beta = 0.1$, and $E_{\rm J}/E_{\rm C} = 50$) results in a YY capacitive coupling strength between qubits 2 and 3 within the USC regime, i.e., $g_{{\rm C}, yy} / \omega_{{\rm q}i} \approx 0.2$. Similarly, for the inductive coupling, setting $E_{\rm J,g}/E_{\rm J} \sim 10^{-2}$, it is possible to achieve USC between qubits 1 and 2 in the XX interaction channel.

\acknowledgments
R.S. and S.S. acknowledges financial support under the National Recovery and Resilience Plan (PNRR), Mission 4, Component 2, Investment 1.4, Call for tender No. 1031 published on 17/06/2022 by the Italian Ministry of University and Research (MUR), funded by the European Union – NextGenerationEU, Project Title “National Centre for HPC, Big Data and Quantum Computing (HPC)” – Code National Center CN00000013 – CUP D43C22001240001.
R.S. and S.S. acknowledge support by the Army Research Office (ARO)
through Grant No. W911NF1910065.
F.N. is supported in part by: the Japan Science and Technology Agency (JST) [via the CREST Quantum Frontiers program Grant No. JPMJCR24I2, the Quantum Leap Flagship Program (Q-LEAP), and the Moonshot R\&D Grant Number JPMJMS2061], and the Office of Naval
Research (ONR) Global (via Grant No. N62909-23-1-2074). 
A.M. was supported by the Polish National Science Centre (NCN) under the Maestro Grant No. DEC-2019/34/A/ST2/00081.

\bibliography{XYspinchain.bib}
\end{document}